\documentclass[prd,
aps,
nofootinbib,nobibnotes,
preprint,
]{revtex4-1}

\usepackage{graphicx}
\usepackage{dcolumn}
\usepackage{bm}
\usepackage{amsmath}
\usepackage{amssymb}
\usepackage{hyperref}

\newcommand{\be}{\begin{equation}}
\newcommand{\ee}{\end{equation}}
\newcommand{\bear}{\begin{eqnarray}}
\newcommand{\eear}{\end{eqnarray}}

\begin{document}

\title{Thermalization of Quark-Gluon Plasma in Magnetic Field \\ at Strong Coupling}

\preprint{RBRC-1133}
\author{ Kiminad A. Mamo }
 \affiliation{Department of Physics, University of Illinois, Chicago, Illinois 60607, USA}
\author { Ho-Ung Yee }
 \affiliation{Department of Physics, University of Illinois, Chicago, Illinois 60607, USA}
 \affiliation{RIKEN-BNL Research Center, Brookhaven National Laboratory, Upton, New York
11973-5000}

\begin{abstract}

We study thermalization of strongly coupled gauge theory plasma in the presence of magnetic field using the AdS/CFT correspondence. We utilize the falling energy-shell model as a holographic description of gauge theory plasma undergoing thermalization, and find the effect of magnetic field on thermalization time in various space-time dimensions. Our results demonstrate that magnetic field universally hastens thermalization of strongly coupled gauge theory plasma.
		
\end{abstract}

\maketitle

\section{Introduction} Recently, studying the effects of magnetic field on the properties of quark-gluon plasma (QGP) has received a considerable amount of attention, as the QGP created in ultra-relativistic heavy-ion collisions may be subject to a huge coherent magnetic field produced by many spectator nucleons~\cite{Kharzeev:2007jp}. Magnetic field can probe several key dynamical~\cite{Bali:2011qj,Mamo:2015dea} and topological~\cite{Fukushima:2008xe,Son:2004tq,Kharzeev:2010gd} properties of QGP related to flavor symmetry of QCD, which are not easily accessible by other gluonic observables, providing important complementary characterization of the created QGP.
In this paper, we study the effect of external magnetic field on the thermalization of QGP
at strong coupling regime using the AdS/CFT correspondence~\cite{Maldacena:1997re}. Magnetic field can potentially be important in the thermalization of QGP in heavy-ion collisions, since the thermalization occurs at an early stage of heavy-ion collisions when the magnetic field is strong, before it dies out with time.

In our study, the QGP that undergoes thermalization is modeled by falling of a thin spatial mass shell to the bottom of AdS space, forming a black-hole at the end of thermalization~\cite{Lin:2008rw}. See Refs.~\cite{Bhattacharyya:2009uu,Chesler:2010bi,Heller:2012km,Wu:2011yd,Garfinkle:2011hm,Garfinkle:2011tc} for other approaches. We treat the magnetic field as external, and use known solutions of AdS geometries with magnetic field at zero and finite temperatures.
In a thin-shell approximation, we join two static solutions, one with zero temperature and the other with finite temperature, across the falling shell via the Israel junction condition~\cite{Israel:1966rt}. At each time, the location of the shell in the energy coordinate (holographic coordinate) divides the AdS space into two regions: one with the geometry of zero temperature that is not yet thermalized, and the other with finite temperature that is thermalized. As the shell falls down towards infrared, eventually forming a black-hole, the AdS space becomes filled with the geometry with finite temperature, representing dynamical thermalization. The proper time (or Eddington-Finkelstein time) by which the shell forms a black-hole can be a reasonable definition of thermalization time in the model~\cite{Lin:2013sga}.

\section{AdS geometries with zero and finite temperatures}
For joining of two static solutions across a thin falling shell to work, each static solution one uses for the two different regions that the shell divides the space-time into, must be isotropic and homogeneous: this requirement is seen in the Israel junction condition in the subsequent analysis. The underlying reason for this requirement can be understood by the Einstein-Maxwell equations with a source (the shell) viewed as an initial value problem. The shell (which is assumed to be neutral) starting from rest at initial time and moving along its trajectory would normally source metric perturbations inside its future light cone while it falls down by its own gravity. There is no a priori reason to expect that the resulting geometry in the future will simply be given by joining of two static geometries across
the falling shell: one instead expects gravitational waves emanating from the shell. In the presence of isotropy and homogeneity however, a powerful uniqueness theorem of Einstein-Maxwell theory dictates that a (neutral) homogeneous and isotropic solution in a connected region with no sources is completely fixed by its conserved energy density, and must take a form of static black-hole with that conserved energy density. The two regions bounded by the shell have constant energy densities differing by the energy density of the shell, and since these energies are conserved, the geometries in each region are fixed by the uniqueness theorem to be those static geometries with conserved energy densities. This is the physical reason why the falling shell ansatz works: simply put, no gravitational radiation is possible in isotropic and homogeneous collapse \cite{thorn}.

We look for AdS geometries with magnetic field which possess isotropy and homogeneity. In dimension $D=5$ (corresponding to 4-dimensional QGP) a single magnetic field necessarily breaks isotropy. To overcome this difficulty, we consider $\mathcal{N}=4$ super-Yang-Mills gauge theory with global $SO(6)_{R}$ R-symmetry
that allows three orthogonal magnetic fields from each $U(1)^3\subset SO(6)_R$ of equal magnitude. Although our model for the magnetic field from R-symmetry of $\mathcal{N}=4$ super Yang-Mills theory has differences from QCD, such as the charge content of the matter fields, we expect that the universal feature we observe in this model could indicate the similar trend in real QCD in strong coupling regime.

The corresponding theory in AdS is the gauged $U(1)^3$ supergravity which is a particular Einstein-Maxwell-Scalar theory. It admits an exact solution with three orthogonal magnetic fields of equal magnitude that ensures isotropy and homogeneity of the energy-momentum tensor \cite{Donos:2011qt}. The action is given by
\begin{align}\label{lag}
(16\pi G_{5})\mathcal{L}&= (R-V)-\frac{1}{2}\,\sum_{I=1}^{2}\,(\partial\phi_{I})^2-\frac{1}{4}\,\sum_{a=1}^{3}X_{a}^{-2}\,(F^{a})^2+\frac{1}{4\sqrt{-g_{5}}}\epsilon^{\mu\nu\rho\sigma\lambda}F_{\mu\nu}^{1}F_{\rho\sigma}^{2}A_{\lambda}^{3}\,,
\end{align}
where $F_{\mu\nu}^a=\partial_{\mu}A_{\nu}^a-\partial_{\nu}A_{\mu}^a$ and
\begin{align}
V&=-\frac{4}{L^2}\,\sum_{a=1}^{3}X_{a}^{-1}\notag\\
X_{1}&=e^{-\frac{1}{\sqrt{6}}\phi_{1}-\frac{1}{\sqrt{2}}\phi_{2}},\quad X_{2}=e^{-\frac{1}{\sqrt{6}}\phi_{1}+\frac{1}{\sqrt{2}}\phi_{2}},\quad X_{3}=e^{\frac{2}{\sqrt{6}}\phi_{1}}\,.
\end{align}
We will set $L=1$ in the following\footnote{More precisely, $L^4= g_{YM}^2 N_c l_s^4$ but the string scale $l_s$ which serves as a UV cutoff cancels in any final field theory observables, and for convenience, we set $L=1$.
In this choice, we have $G_5=\pi/(2N_c^2)$.}.
The field equations derived from the Lagrangian (\ref{lag}) admit an exact magnetically charged AdS$_{5}$ black hole solution \cite{Donos:2011qt}
\begin{align}\label{s5}
ds_{5}^{2}&={dz^2\over f(z) z^2}-{f(z)\over z^2}dt^2 + {\left(d\vec x\right)^2\over z^2}\,,\notag\\
F_{ij}^{a}&=\epsilon^{aij}B\,,\quad \phi_I=0\,,
\end{align}
where $a=1,2,3$ labels three $U(1)$ R-symmetries and $i,j=1,2,3$ are spatial indices. The function $f(z)$ is
\begin{equation}\label{f5}
  f(z)=1-mz^4+\frac{1}{8}B^2z^4\,\log{(mz^4)}\,,
\end{equation}
with $z_{H}=m^{-\frac{1}{4}}$ being the location of the black hole horizon solving $f(z_H)=0$. The parameters $(m,B)$ are related to the temperature $T$ by
\begin{equation}\label{t5}
T=-{f'(z_H)\over 4\pi}=\frac{8m-B^2}{8\pi m^{3/4}}\,.
\end{equation}
Note that at $m=\frac{1}{8}B^2$, the temperature of the black hole (\ref{t5}) goes to zero, and hence the extremal zero temperature solution in the presence of magnetic fields is given by
\begin{equation}\label{f05}
  f_{0}(z)=1-\frac{1}{8}B^2z^4+\frac{1}{8}B^2z^4\,\log{(\frac{1}{8}B^2z^4)}\,.
\end{equation}
We map this to the field theory vacuum in the presence of magnetic field.
We must have $m\ge \frac{1}{8}B^2$ for thermodynamic stability.

In 4-dimensional AdS space (corresponding to a field theory in 3-dimensions), one can realize isotropy and homogeneity with a single magnetic field $F_{12}=B$.
The exact black hole solution with magnetic field in the Einstein-Maxwell theory is known~\cite{Romans:1991nq}\begin{equation}\label{s4}
ds_{4}^{2}={dz^2\over f(z) z^2}-{f(z)\over z^2}dt^2 + {\left(d\vec x\right)^2\over z^2}\,,
\end{equation}
where
\begin{equation}\label{f4}
  f(z)=1-mz^3+B^2z^4\,.
\end{equation}
The location of the black hole horizon is given by $f(z_H)=0$, and the temperature $T$ is
\begin{equation}\label{t4}
T=-{f'(z_H)\over 4\pi}=\frac{3-B^2z_{H}^4}{4\pi z_{H}^3}\,.
\end{equation}
When $m=m_0\equiv\frac{4}{3^{3/4}}B^{3/2}$, the temperature of the black hole solution (\ref{t4}) becomes zero, and hence the extremal zero temperature solution is given by the blackening factor,
\begin{equation}\label{f04}
  f_{0}(z)=1-\frac{4}{3^{3/4}}B^{3/2}z^3+B^2z^4\,.
\end{equation}

The metric (\ref{s4}) with $m=0$, that is $f(z)=1+B^2 z^4$, is a solution of the Einstein-Maxwell equation without black-hole horizon. The reason why it can not be the solution for zero temperature is its violation of causality: the speed of light in the bulk AdS with respect to the field theory coordinates $(t,\vec x)$ at position $z$ is $c(z)=f(z)$, which has to be less than 1 to respect causality of the field theory~\cite{Bak:2004yf}. This means that this geometry should be excluded in a meaningful AdS/CFT correspondence.

 \section{Holographic thermalization with magnetic field}
The thin shell initially starting from rest at a position $z_i=1/\pi Q_s$ collapses
from the UV region of small $z$ to the IR region of large $z$ under its own gravity, eventually passing through its black-hole horizon by which we have thermalization. The geometry is constructed by joining a black hole solution with finite temperature (\ref{f5}) above the shell in the UV region with the zero temperature solution (\ref{f05}) below the shell, across the trajectory of the shell in $(t,z)$ coordinates that is determined by Israel junction conditions.

The metric induced on the 4-dimensional world-volume $\Sigma$ of the shell can be written in a conformal form
\begin{equation}\label{ind}
ds^2_\Sigma ={-d\tau^2 +\left(d\vec x\right)^2\over \left(z(\tau)\right)^2}\,,
\end{equation}
where $z(\tau)$ is the position of the shell in $z$ coordinate at a conformal time $\tau$.
Continuity of the metric across the shell
requires identifying $\vec x$ on $\Sigma$ with $\vec x$ in the background.
The trajectory of the shell with respect to $(t_U,z)$ coordinates in the upper (UV) region of space-time  parameterized by the conformal time $\tau$, that is $\left(t_U(\tau),z(\tau)\right)$,
determines the induced metric on $\Sigma$. Comparing time component of that with (\ref{ind}) gives
\begin{equation}\label{cond1}
f\left(z(\tau)\right)\dot t_U^2(\tau)-{\dot z^2(\tau)\over f\left(z(\tau)\right)}=1\,,
\end{equation}
where $\cdot\equiv{d\over d\tau}$. This relates $t_U$ and $\tau$, given a trajectory $z(\tau)$.
Similarly, the same trajectory with respect to the IR coordinates $(t_L(\tau),z(\tau))$
should satisfy the condition
\begin{equation}\label{cond2}
f_{0}\left(z(\tau)\right)\dot t_L^2(\tau)-{\dot z^2(\tau)\over f_{0}\left(z(\tau)\right)}=1\,,
\end{equation}
that gives a relation between $t_L$ and $\tau$
once the trajectory $z(\tau)$ is found. 
Finally the Israel junction condition is
\be
\left[K_{ij}-\gamma_{ij}K\right]=-8\pi G_5 S_{ij}\,,
\ee
where $[A]\equiv A_L-A_U$, $S_{ij}$ is the energy-momentum tensor on the shell,
and $\gamma_{ij}$ is the induced metric on the shell with respect to the shell coordinate $\xi^i=(\tau,\vec x)$. The $K_{ij}^{U/L}$ are extrinsic curvatures evaluated on the shell from the upper and lower regions respectively,
\be
K_{ij}={\partial x^\alpha\over\partial\xi^i}{\partial x^\beta\over\partial \xi^j} \nabla_\alpha n_\beta=-n_\alpha\left({\partial^2 x^\alpha\over\partial \xi^i\partial\xi^j}+\Gamma^\alpha_{\beta\gamma} {\partial x^\beta\over\partial\xi^i}{\partial x^\gamma\over\partial\xi^j}\right)\,,
\ee
with the unit normal vectors $n^\mu_{L/R}$ to the surface $\Sigma$ pointing to the direction of increasing $z$.
They are given by 
\bear
n_U&=&\left({z\dot z\over f(z)}\right){\partial\over\partial t}+\left(zf(z)\dot t\right){\partial\over\partial z}\,,\nonumber\\
n_L&=&\left({z\dot z\over f_{0}(z)}\right){\partial\over\partial t}+\left(zf_{0}(z)\dot t\right){\partial\over\partial z}\,,
\eear
where all quantities are evaluated on the shell.
The non-vanishing components of $K_{ij}^{U/L}$ are 
\bear
K^U_{\tau\tau}&=&-{\dot t_U\over z}\left({f\left(f'+2\ddot z\right)\over 2\left(f+\dot z^2\right)}-{f\over z}\right)\,,\nonumber\\
K^U_{ij}&=&-{\dot t_U f\over z^2} \delta_{ij}\,,\quad i,j=1,2,3\,,\nonumber\\
K^L_{\tau\tau}&=&-{\dot t_L\over z}\left({f_{0}\left(f_{0}'+2\ddot z\right)\over 2\left(f_{0}+\dot z^2\right)}-{f_{0}\over z}\right)\,,\nonumber\\
K^L_{ij}&=&-{\dot t_L f_{0}\over z^2} \delta_{ij}\,,\quad i,j=1,2,3\,.
\eear
where $'\equiv {d\over dz}$. Assuming energy-momentum tensor on the shell of a conformal form,
\be
S_{ij}=4p(z)u_i u_j +\gamma_{ij} p(z)\,,\quad u_i=\left({1\over z},0,0,0\right)\,,
\ee
with the pressure $p(z)$ to be determined, the junction condition becomes 
\bear
f_{0}\dot t_L-f \dot t_U&=&8\pi G_5 p(z)\,,\nonumber\\
\dot t_L{zf_{0}\left({f_{0}'\over 2}+\ddot z\right)\over (f_{0}+\dot z^2)}-\dot t_U{zf\left({f'\over 2}+\ddot z\right)\over (f+\dot z^2)}&=&4\cdot8\pi G_5 p(z)\,.
\eear
Removing $p(z)$ from the above equations and using
\be
\dot t_L={\sqrt{f_{0}+\dot z^2}\over f_{0}}\,,\quad \dot t_U={\sqrt{f+\dot z^2}\over f}\,,
\ee
from (\ref{cond1}) and (\ref{cond2}), the resulting equation for $\dot z$ is integrable to give
\be
\dot z=\sqrt{\left({Cz^4\over 2}+{f_{0}(z)-f(z)\over 2Cz^4}\right)^2-f_{0}(z)}\,,\label{dotz1}
\ee
with a constant of motion $C>0$. This reproduces the one in Ref.~\cite{Lin:2013sga}.
We choose to express the falling trajectory in terms of the boundary time $t_U$ which can be identified with the field theory (QCD) time on the boundary. Using the relation (\ref{cond1}), the solution (\ref{dotz1}) translates to
\be
{dz\over dt_U}=f(z)\sqrt{{\left({Cz^4\over 2}+{f_{0}(z)-f(z)\over 2Cz^4}\right)^2-f_{0}(z)\over
\left({Cz^4\over 2}+{f_{0}(z)-f(z)\over 2Cz^4}\right)^2-(f_{0}(z)-f(z))}}\,,\label{dotz2}
\ee
which we solve numerically. More precisely, the thermalization time is defined as the Eddington-Finkelstein time
when the mass shell passes through its black-hole horizon~\cite{Lin:2013sga}.

\begin{table}[t]
\begin{center}
\begin{tabular}{c|c|c|c}
\hline $B$ (${\rm fm}^{-2}$)& $z_H$ (fm) &$m$ (${\rm fm}^{-4}$) & $C$ (${\rm fm}^{-4}$)\\\hline
0&0.209&525.5&263.7\\
2.06&0.209&527.6&262.6\\
5.16&0.208&538.7&260.2\\
25.81&0.187&810.6&272.1\\
51.61&0.162&1469.1&331.0\\\hline
\end{tabular}
\caption{(AdS$_{5}$) Parameters of our numerical solutions for RHIC with a late-time temperature $T=300$ MeV and several exemplar values of $B=0;\,0.08~\rm GeV^2;\,0.2~\rm GeV^2;\,1~\rm GeV^2;\,2~\rm GeV^2$. }\label{tab1}
\end{center}
 \end{table}

\begin{figure}[t]
\begin{center}
\begin{tabular}{cc}
\includegraphics[width=7cm]{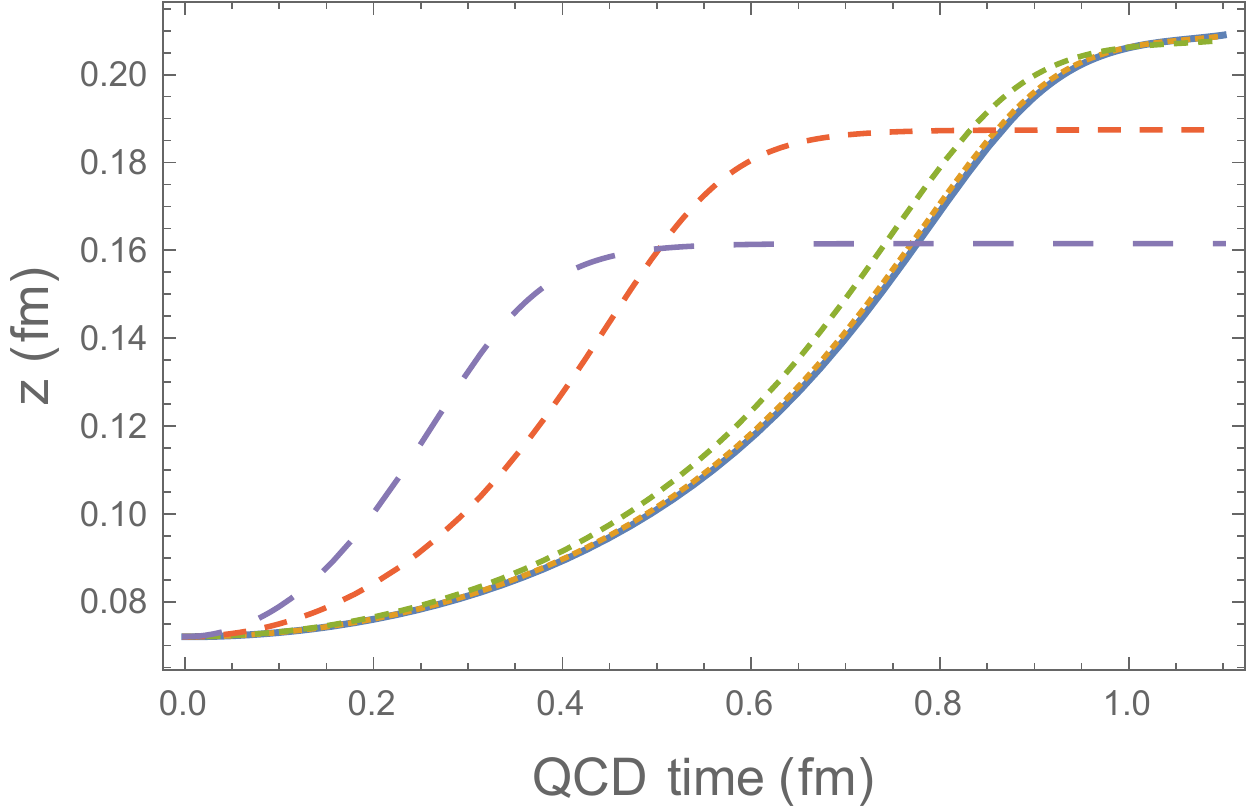}

&
\includegraphics[width=7cm]{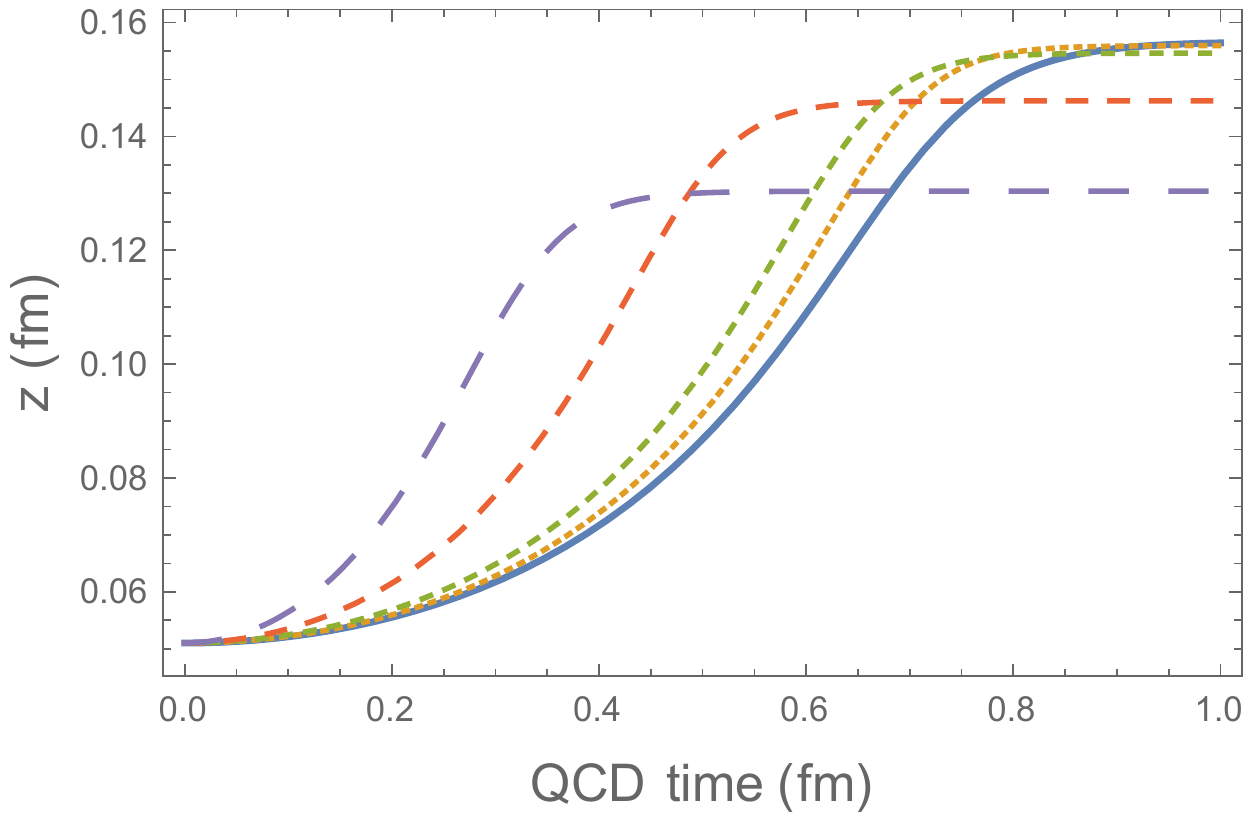}

\\
(a) & (b)
\end{tabular}
\caption{(AdS$_{5}$) Thermalization history of falling mass shell for RHIC (left) and LHC (right).
		The late-time temperature is fixed to be $T=300\, (400)$ MeV for RHIC (LHC), and the magnetic fields are $B=0\,(0)~(\rm solid~blue);~0.08\,(0.3)~(\rm orange);~0.2\,(0.52)~(\rm green);~1\,(1.32)~(\rm red);~2\,(2.64)~(\rm violet)$ $\rm GeV^2$ for RHIC (LHC). Thermalization time is when the curve reaches its plateau at the horizon. \label{fig1}}
\end{center}
\end{figure}

Following \cite{Lin:2013sga}, we set our initial condition of the falling mass shell in terms of the saturation scale $Q_s$, which governs the initial gluon distribution, as
\be
z\left(t_U=0\right)=z_i={1\over \pi Q_s}\,,\quad \dot z\left(t_U=0\right)=0\,.
\ee
We measure $z$ in units of fm.
For RHIC, we take $Q_s=0.87$ GeV~=~4.42 ${\rm fm}^{-1}$, and for LHC we have $Q_s=1.23$ GeV~=~6.24 ${\rm fm}^{-1}$. In Table \ref{tab1}, we show parameters of our numerical solutions after fixing the final thermalization temperature to be $T=300$ MeV for RHIC for several exemplar values of magnetic field.
In Figure \ref{fig1}, we show the time history of falling mass shell in the field theory (QCD) time $t_U$ for a few exemplar values of magnetic field $B$ with a fixed final temperature. The plots clearly indicate that the presence of magnetic field speeds up the thermalization of the plasma: the stronger the magnetic field, the shorter the thermalization time.
More precisely, the thermalization time is defined as the Eddington-Finkelstein time
when the mass shell passes through its black-hole horizon~\cite{Lin:2013sga}. However, it is qualitatively similar to the time in Schwarz coordinate $t_U$ we show when the mass shell falls close to the horizon.

\begin{figure}
  \centering
  \includegraphics[width=7cm]{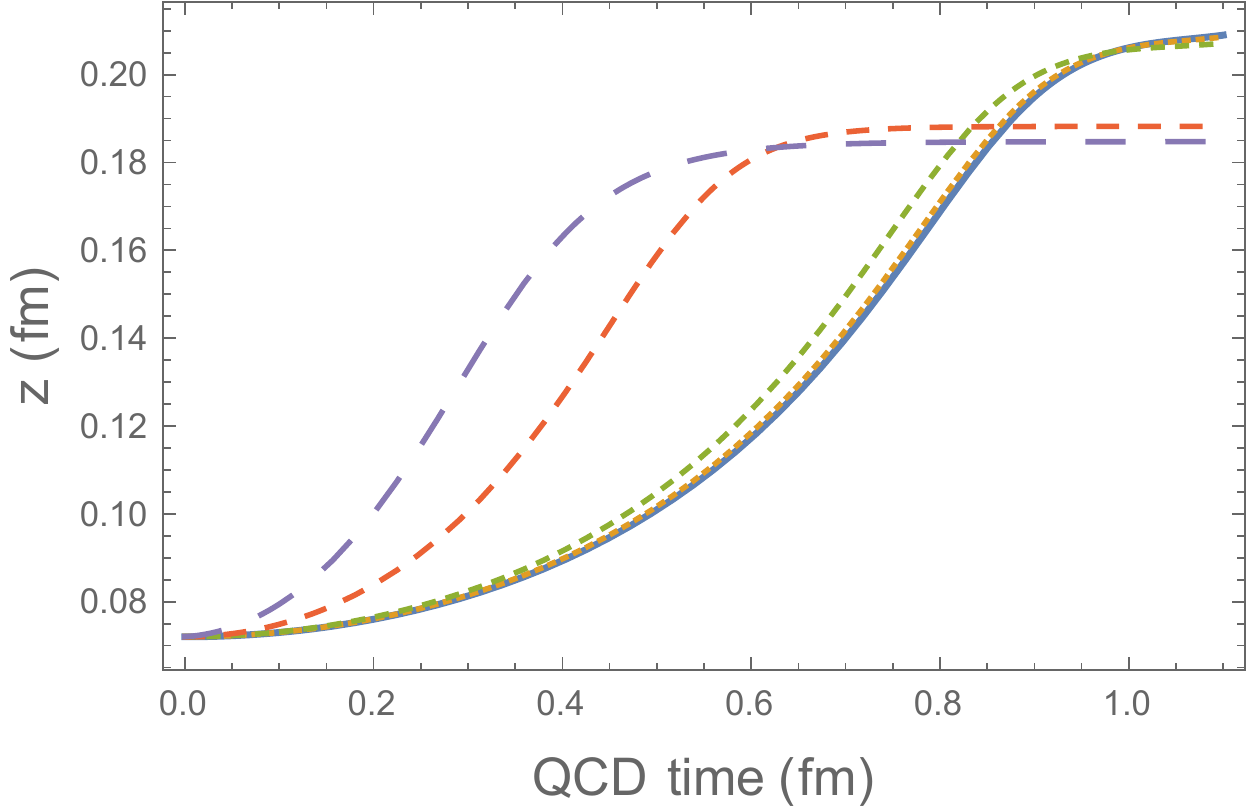}
  \caption{(AdS$_{5}$) Thermalization history of falling mass shell for fixed energy density, and varying magnetic field $B~=~0~(\rm solid~blue);~0.08~(\rm orange);~0.2~(\rm green);~1~(\rm red);~2~(\rm violet)$ $\rm GeV^2$ for RHIC.}\label{newfig}
\end{figure}

Instead of fixing final temperature, we also study the case where the energy density measured from zero temperature but finite $B$ state is fixed while we vary magnetic field, that is we fix
$\Delta \epsilon \equiv \epsilon(T,B)-\epsilon(T=0,B)$, which can be interpreted as the energy density thrown by colliding nuclei into the background magnetic field.
Explicitly, we have
\be\label{energy}
\Delta\epsilon=\frac{N_{c}^2}{4\pi^2}\left({3\over 2}m+{3\over 16}B^2\left(\log\left(B^2\over 8m\right)-1\right)\right)\,,
\ee
which determines the parameter $m$ in the geometry, given a fixed $\Delta\epsilon$ and varying $B$. In Figure \ref{newfig}, we show the resulting time trajectories of energy shell with $\Delta\epsilon$
chosen to be the energy density of $T=300$ MeV, $B=0$ state. Our observation of faster thermalization with magnetic field seems robust.

\begin{figure}[t]
	\centering
	\includegraphics[width=7cm]{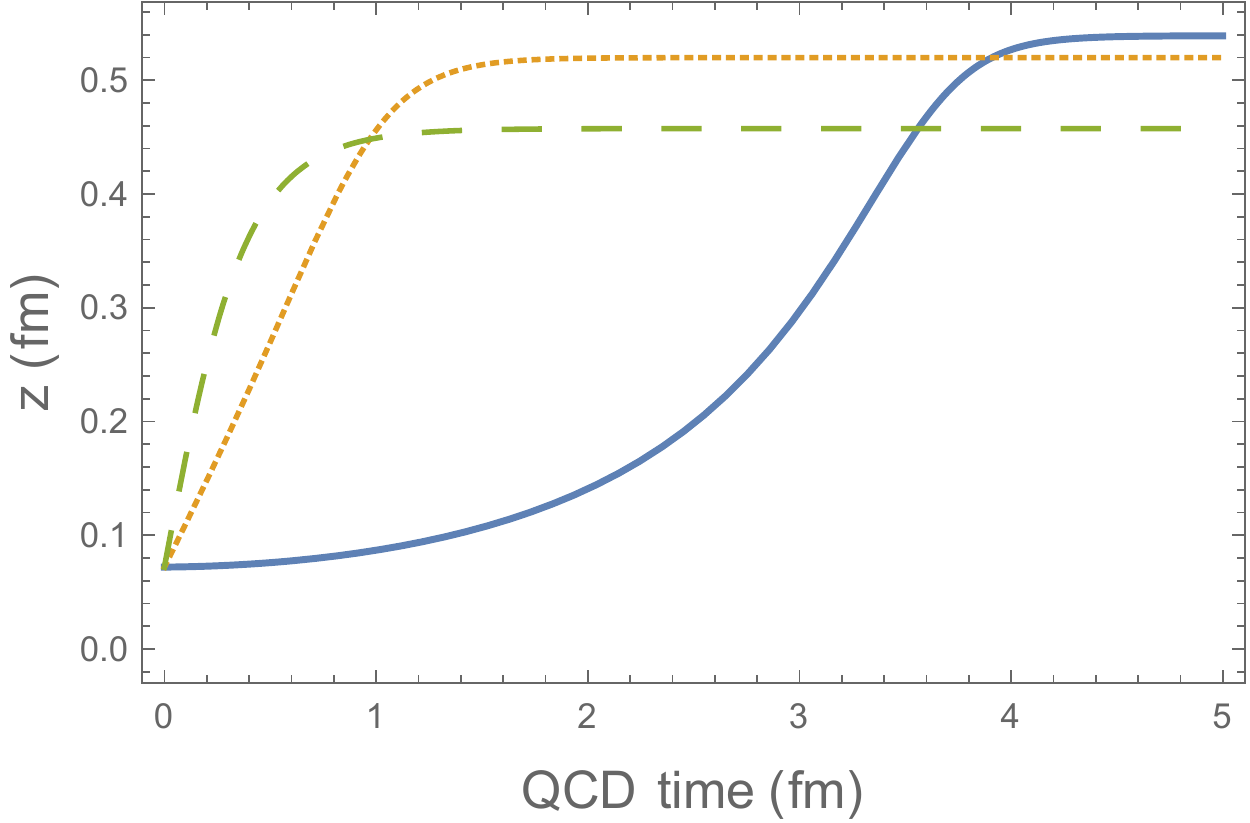}
		\caption{(AdS$_{4}$) Thermalization history of falling mass shell in $AdS_4$ for a late-time temperature of $T=300$, and the magnetic fields are $B=0 ~(\rm solid~blue);~0.08~(\rm orange);~0.2~(\rm green)$ $\rm GeV^2$. \label{fig2}}
\end{figure}

To examine whether our conclusion depends on the number of dimensions the field theory resides in, we study the thermalization of plasma in magnetic field in one less dimension.
In $AdS_{4}$ (corresponding to 3-dimensional field theory), the analysis is the same with (\ref{f4}) in the place of (\ref{f5}), and (\ref{f04}) in the place of (\ref{f05}), but with the energy-momentum tensor on the shell taking a 3-dimensional conformal form,
\be
S_{ij}=3p(z)u_i u_j +\gamma_{ij} p(z)\,,\quad u_i=\left({1\over z},0,0\right)\,.
\ee
We arrive at
\be
{dz\over dt_U}=f(z)\sqrt{{\left({Cz^3\over 2}+{f_{0}(z)-f(z)\over 2Cz^3}\right)^2-f_{0}(z)\over
\left({Cz^3\over 2}+{f_{0}(z)-f(z)\over 2Cz^3}\right)^2-(f_{0}(z)-f(z))}}\,,\label{dotz24}
\ee
which can be solved numerically given the constant $C$ which, as before, should be determined from initial conditions. 
In Figure \ref{fig2}, we show the time history of falling mass shell trajectory in field theory time $t_U$ for a few exemplar values of magnetic field $B$ for 3-dimensional gauge theory with a fixed final temperature. Again, the plots clearly demonstrate that the presence of magnetic field hastens the thermalization.

\section{Summary and discussion}
In the framework of AdS/CFT correspondence, we have studied the thermalization of strongly coupled gauge theory plasma in the presence of magnetic field, utilizing simplified picture of thermalization as falling of a thin homogeneous energy-shell towards the black-hole horizon. Our results in various dimensions have revealed that magnetic field universally hastens thermalization in strong coupling regime. At weak coupling, a strong magnetic field causes the dimensional reduction of the system into 1-dimensional one with lowest Landau levels, and one may study the effects of magnetic field to thermalization at weak coupling in this context. It would be interesting to see how weak coupling result compares with our conclusion in this work at strong coupling.

{\it Note added:} while our work was near its final stage, there appeared~\cite{Fuini:2015hba} which addresses a similar question in a different set-up. Our numerical results are in qualitative
agreement with theirs: for the values of magnetic field relevant at RHIC and LHC, the effect of magnetic field to thermalization time is less significant. Our work is about the universal trend of strong magnetic field hastening plasma thermalization.

\section*{Acknowledgements}
We thank Shu Lin for helpful discussions.

\bibliographystyle{prsty}
\bibliography{ar,tft,qft,books}

\end{document}